\documentclass[letterpaper,11pt]{article}

\usepackage{scrextend}
\usepackage{amsmath,amssymb,epsfig,bbm}
\usepackage[active]{srcltx}
\usepackage[all]{xypic}
\usepackage{MnSymbol}
\usepackage{slashed}
\usepackage{amsthm}

\usepackage{float}

\usepackage{color}

\usepackage{dsfont}

\definecolor{co}{cmyk}{0,0.7,0.3,0}
\definecolor{darkgreen}{cmyk}{1,0,1,.2}
\definecolor{m}{rgb}{1,0.1,1}

\newcommand{\be}{\begin{equation}}
\newcommand{\ba}{\begin{eqnarray}}
\newcommand{\ea}{\end{eqnarray}}
\newcommand{\nn}{\nonumber}

\def\d{\delta}

\def\m{\mu}
\def\n{\nu}
\def\oo{\omega}

\def\OO{\Omega}

\def\ca{{\cal A}}
\def\cb{{\cal B}}

\def\cd{{\cal D}}

\def\cf{{\cal F}}

\def\ch{{\cal H}}

\def\cn{{\cal N}}
\def\co{{\cal O}}

\makeatletter
\newcommand{\eqnum}{\refstepcounter{equation}\textup{\tagform@{\theequation}}}
\makeatother

\newcommand{\pa}{\partial}

\newcommand{\C}{{\Bbb C}}

\newtheorem{thm}{Theorem}[subsection]

\newtheorem{conj}[thm]{Conjecture}
\newtheorem{definition}[thm]{Definition}
\newtheorem{proposition}[thm]{Proposition}
\newtheorem{remark}{Remark}

\newcommand{\bbR}{{\Bbb R}}

\newtheorem*{definition*}{Definition}

\newcommand{\R}{\operatorname{\bold R}}

\fontfamily{yfrak}

\begin{document}

\vskip 25mm

\begin{center}

{\large\bfseries

Non-perturbative Quantum Field Theory  \\
and the Geometry of Functional Spaces

}

\vskip 6ex

Johannes \textsc{Aastrup}$^{a}$\footnote{email: \texttt{aastrup@math.uni-hannover.de}} \&
Jesper M\o ller \textsc{Grimstrup}$^{b}$\footnote{email: \texttt{jesper.grimstrup@gmail.com}}\\ 
\vskip 3ex

$^{a}\,$\textit{Mathematisches Institut, Universit\"at Hannover, \\ Welfengarten 1, 
D-30167 Hannover, Germany.}
\\[3ex]
$^{b}\,$\textit{Independent researcher, Copenhagen, Denmark.}
\\[3ex]

{\footnotesize\it This work is financially supported by Ilyas Khan,  St.\\ EdmundÕs College, Cambridge, United Kingdom and by\\ \vspace{-0,1cm}Regnestuen Haukohl \& K\o ppen, Copenhagen, Denmark.}

\end{center}

\vskip 3ex

\begin{abstract}

In this paper we construct a non-commutative geometry over a configuration space of gauge connections and show that it gives rise to a candidate for an interacting, non-perturbative quantum gauge theory coupled to a fermionic field on a curved background. 
The non-commutative geometry is given by an infinite-dimensional Bott-Dirac type operator, whose square gives the Hamilton operator, and which interacts with an algebra generated by holonomy-diffeomorphisms. The Bott-Dirac operator and the associated Hilbert space relies on a metric on the configuration space of connections, which effectively works as a covariant ultra-violet regulator. We show that the construction coincides with perturbative quantum field theory in a local limit. Questions concerning Lorentz invariance and the fermionic sector as well as the issue of existence are left open.

\end{abstract}

\newpage
\tableofcontents

\section{Introduction}
\setcounter{footnote}{0}

One of the most important questions in contemporary theoretical physics is the rigorous understanding of quantum field theory. This question has two legs; first there is the task of formulating a well-defined non-perturbative framework and secondly there is the task of finding conceptual explanations for the central structures found in quantum field theory. Embedded within these questions lies also the question about what role the gravitational field plays in a quantum framework.
 
In this paper we propose a novel answer to these questions. We show that a candidate for a $3+1$-dimensional interacting, non-perturbative quantum field theory, which includes a gauge field coupled to a fermionic field  on a curved background, emerges from a geometrical construction {\it over} a configuration space of gauge connections. This geometrical construction provides new conceptual interpretations of central ingredients in quantum field theory, such as the canonical commutation relations, the role of the fermionic sector and the geometrical origin of the Hamilton operator itself. Moreover, within this framework general relativity remains essentially classical; the Planck scale screening, which is usually ascribed to a quantum theory of gravity, emerges in this framework as a consequence of representation theory. \\

The non-commutative geometry, that we find, is composed of two basic constituents, namely an algebra of holonomy-diffeomorphisms and an infinite-dimensional Bott-Dirac operator. Holonomy-diffeomorphisms, which encode how tensor degrees of freedom are transported along flows of vector fields, form a natural, non-commutative algebra of functions on a configuration space of gauge connections.  Combined with translations on the underlying configuration space this algebra encodes the canonical commutation relations of a quantum gauge theory \cite{Aastrup:2014ppa}. The Bott-Dirac operator then combines all infinitesimal translations on this configuration space into a single operator that effectively encodes metric information on the infinite-dimensional space.

The setup presented in this paper is similar to the construction proposed in \cite{Aastrup:2017atr}. There, a similar Bott-Dirac operator was shown to produce the free Hamiltonian of a Yang-Mills theory coupled to a fermionic field. The problem with that construction was, however, that it did not preserve the gauge symmetry; both the Bott-Dirac operator and the Hilbert space representation involved a broken gauge symmetry. 
In the present setup these problems are solved. The square of the new Bott-Dirac operator produces the {\it full} Hamilton operator of an interacting Yang-Mills system coupled to a fermionic field and the Hilbert space representation of both the Bott-Dirac operator and the algebra of holonomy-diffeomorphisms, does not break the gauge symmetry in the previous indiscriminate manner.

Hilbert space representations of the $\mathbf{QHD}(M)$ algebra, which is the algebra generated by holonomy-diffeomorphisms and translation on the underlying configuration space, were first constructed in \cite{Aastrup:2017vrm}. The representations found there rely on a Sobolev norm, whose role it is to dampen ultra-violet degrees of freedom; i.e the Sobolev norm work as an ultra-violet regularisation, and since this Sobolev norm involve a Hodge-Laplace operator, it automatically breaks the gauge symmetry. 
In the present setup this problem is solved by constructing the Sobolev norm in a covariant manner. What this means, basically, is that we introduce a gauge-covariant UV regularisation. When this is done the Sobolev norm is promoted to a gauge-invariant metric on the configuration space of gauge connections. This metric then becomes the cornerstone in our construction, both when it comes to formulate the Hilbert space representation and when it comes to define the Bott-Dirac operator.
Hence, this gauge-covariant regularisation is to be viewed as a part of the theory and not merely as a computational artifact.

It is widely believed that distances shorter than the Planck scale are operational meaningless -- simple arguments combining general relativity and quantum mechanics strongly supports this view \cite{Doplicher:1994tu} -- and this ultra-violet screening is mostly thought to be a consequence of a theory of quantum gravity, where a quantized metric will, in one way or another, have a discrete spectrum and hence preclude arbitrarily short distances.
What we suggest is that the ultra-violet screening may in fact be produced by quantum field theory itself as a consequence of the representation theory of basic algebra of observables such as the $\mathbf{QHD}(M)$ algebra.
If this were to be the case then gravity need no quantization.


Thus, what we propose is essentially a gravitational theory {\it over} a configuration space of connections, and what we find is that {\it if} such a theory exist, then it will effectively look like a quantum field theory on a curved background. 

Concerning existence then we showed in \cite{Aastrup:2017vrm} that Hilbert space representations exist in the specific case where the UV regularisation is {\it not} covariant. In the covariant case we do not have a proof of existence. We do, however, list a number of indications suggesting that the required convergence can also be obtained in the covariant case.

With respect to the metric structure then the notion of a distance on a configuration space of gauge connections is not new but
was discussed already by Feynman \cite{Feynman:1981ss} and Singer \cite{Singer:1981xw} (see also \cite{Orland:1996hm} and references therein). The construction which we propose is, however, different from what has previously been discussed.

Quantum field theory is usually founded on the two basic principles of locality and Lorentz invariance. In the axiomatic approaches, for instance, these principles are encoded in the Osterwalder Schrader axioms \cite{Osterwalder:1973dx} for the Euclidean theory and in the G\.{a}rding-Wightman \cite{Wightman} or the Haag-Kastler \cite{Haag:1963dh} axioms for the Lorentzian theory. The quantum field theory that we propose will, however, be inherently non-local due to the covariant regularisation. Effectively this means that the theory involves an infinite series of higher order derivative terms. 
Whether the theory will be Lorentz invariant is at the moment not known. Certain technicalities, such as the occurrence of vector-fermions, suggest that it will not be the case. It is, however, possible that these occurrences are simply an indication that we have not yet fully understood the representation theory of the infinite-dimensional Clifford algebra, that is used to construct the Bott-Dirac operator. On the other hand, there exist the possibility that the covariant UV regularisation will break the Lorentz symmetry at the Planck scale, something that may very well be within experiemental bounds \cite{Jacobson:2004rj}. In that case the Lorentz symmetry will likely be amended with a scale transformation.

As already mentioned, the geometrical construction, that we find, may shed a new light on the fermionic sector in quantum field theory. Whereas the canonical commutation relations (CCR) algebra in our construction emerge from the $\mathbf{QHD}(M)$ algebra the canonical anti-commutation relations (CAR) algebra emerge from the construction of the infinite-dimensional Bott-Dirac operator just like the ordinary, finite-dimensional Clifford algebra emerges from a Dirac operator on a spin-manifold. The CAR algebra is here simply the infinite-dimensional Clifford algebra associated to an infinite-dimensional Dirac operator. In this interpretation the fermionic sector is therefore seen as an integral part of a geometrical construction on the underlying space of field configurations.



\subsection{Outline of the central idea}

Let us begin with a rough outline of the central idea behind our construction.
We start with a configuration space $\ca$ of gauge connections and consider translations thereon. Two arbitrary connections $A$ and $A'$ always differ by a one-form $\omega$
$$
A' = A+\omega,
$$
which corresponds to a translation operator $U_\omega$
$$
U_\omega \xi(A) = \xi(A+\omega)
$$
on functions $\xi$ on $\ca$. 
Now, if we consider infinitesimal translations $\frac{\pa}{\pa A_i}$, where $\{A_i\}$ is an orthonormal basis of one-forms, then we can form a Bott-Dirac operator on $\ca$ of the form 
$$
B= \sum_{i=1}^\infty c_i \frac{\pa}{\pa A_i} + \bar{c}_i F_i
$$
where $F_i$ is the curvature of $A_i$ and where $(c_i,\bar{c}_i)$ are elements in an infinite-dimensional Clifford algebra. The square of $B$ gives us then the Hamiltonian of a Yang-Mills theory coupled to a fermionic field
$$
B^2 = \sum_{i=1}^\infty  \left(\frac{\pa}{\pa A_i}\right)^2 +  \left( F_i\right)^2 + "\mbox{fermionic terms}"
$$
in a form, which resembles an infinite-dimensional harmonic oscillator. Here the fermionic sector emerges from the infinite-dimensional Clifford algebra, which is required in order to construct $B$. 

The aim of this paper is to rigorously develop this idea, i.e. to construct a non-commutative geometry on the configuration space $\ca$ constituted by an algebra of holonomy-diffeomorphisms, the Bott-Dirac operator $B$ and a Hilbert space representation hereof, and to use this geometry to derive a candidate for a non-perturbative quantum field theory involving both bosonic and fermionic degrees of freedom on a curved background.

\subsection{Outline of the paper}

We begin in section \ref{sektion2} with the definitions of the $\mathbf{HD}(M)$ and $\mathbf{QHD}(M)$ algebras and show that these algebras encode the canonical commutation relations of a gauge theory. In section \ref{sektion_metrik_paa_a} we then define a metric on a configuration space of gauge connections and use this in section \ref{secdirac} to define Dirac and Bott-Dirac operators on this configuration space. We then construct a Hilbert space in section \ref{sechilbert} and conjecture that it carries a representation of the $\mathbf{QHD}(M)$ algebra as well as the Bott-Dirac operator. Section \ref{secCS} is concerned with the construction of the ground state, which is to lie in the kernel of the Bott-Dirac operator. This construction is then, in section
\ref{conqft}, shown to produce the Hamilton operator of a quantized gauge field coupled to a fermionic field in a local limit. Finally we discuss the general case with a Levi-Civita connection in $T\ca$ in section 8 and end with a discussion in section 9.

\vspace{0.5cm}
\noindent{\bf Notation}\vspace{0.3cm}

Throughout the paper $M$ denotes a spatial manifold, which is 3-dimensional and compact unless otherwise stated. Throughout the paper we denote by $g$ a fixed metric on $M$. We denote by $\{{\bf m},{\bf m}_1,{\bf m}_2 \ldots\}$ points in $M$ and by $\{m_\m\}$ a corresponding coordinate system where $\{\m,\n, \ldots\}$ are spatial indices. We denote by $(x_1,x_2,\ldots)$ coordinates in $\R^n$. Furthermore, we use $\{a,b,\ldots\}$ to index Lie-algebras. Finally, $\{i,j,\ldots\}$ label orthonormal bases of $\OO^1(M,\mathfrak{g})$ and $L^2(M,\mathfrak{g})$ where $\mathfrak{g}$ is a Lie algebra.

\section{The $\mathbf{HD}(M)$ and $\mathbf{QHD}(M)$ algebras}
\label{sektion2}

In this section we introduce the algebras $\mathbf{HD}(M)$ and $\mathbf{QHD}(M)$, which are generated by parallel transports along flows of vector-fields and for the latter part also by translation operators on an underlying configuration space of connections. We also show that the $\mathbf{QHD}(M)$ algebra encode the canonical commutation relations of a gauge theory. The $\mathbf{HD}(M)$ algebra was first defined in \cite{Aastrup:2012vq,AGnew} and the $\mathbf{QHD}(M)$ algebra in \cite{Aastrup:2014ppa}.

\subsection{The holonomy-diffeomorphism algebra}
\label{beent}

Let $M$ be a compact smooth 3-dimensional manifold, let $G$ be a compact Lie group, and let $\upsilon :G\to M_j (\mathbb{C})$ be a unitary faithful representation. Consider the vector bundle $S=M\times \C^j$ over $M$ as well as the space of $G$ connections acting on the bundle. Given a metric $g$ on $M$ we get the Hilbert space $L^2(M,S,dg)$, where we equip $S$ with the standard inner product. Given a diffeomorphism $\phi:M\to M$ we get a unitary operator $\phi^*$ on  $L^2(M,S,dg)$ via
$$( \phi^* (\xi ))(\phi ({\bf m}) )= (\Delta \phi )(M)  \xi ({\bf m}) , $$
where  $\Delta \phi ({\bf m})$ is the volume of the volume element in $\phi ({\bf m})$ induced by a unit volume element in $ {\bf m}\in M$ under $\phi $.      

Let $X$ be a vector field on $M$ and let $A$ be a $G$-connection acting on $S$.  Denote by $t\to \exp_t(X)$ the corresponding flow. Given ${\bf m}\in M$ let $\gamma$ be the curve  
$$\gamma (t)=\exp_{t} (X) ({\bf m}) $$
running from ${\bf m}$ to $\exp_1 (X)({\bf m})$. We define the operator 
$$e^X_A :L^2 (M , S, dg) \to L^2 (M ,  S , dg)$$
in the following way:
we consider an element $\xi \in L^2 (M ,  S, dg)$ as a $\C^j$-valued function, and define 
\begin{equation}
  (e^X_A \xi )(\exp_1(X) ({\bf m}))=  ((\Delta \exp_1) (m))  \hbox{Hol}(\gamma, A) \xi ({\bf m})   ,
  \label{chopin1}
 \end{equation}
where $\hbox{Hol}(\gamma, A)$ denotes the holonomy of $A$ along $\gamma$. Note that $e^X$ is a unitary operator. Again, the factor $(\Delta \exp_1) (M)$ is accounting for the change in volumes, rendering $e^X_\nabla$ unitary.  

Let $\ca$ be the space of $G$-connections acting on $S$. We have an operator valued function on $\ca$ defined via 
\begin{equation}
\ca \ni A \to e^X_A  . 
\nn
\end{equation}
We denote this function $e^X$.  We call this operator a holonomy-diffeomorphism\footnote{In \cite{AGnew} the definition of a holonomy-diffeomorphisms also included a function $f\in C^\infty_c (M)$, which gives another operator valued function $fe^X$ on $\ca$. For reasons that were given in \cite{Aastrup:2018coc} such local holonomy-diffeomorphims are not viable in our present construction.}. 
Denote by $\cf (\ca , \mathbb{B} (L^2(M, S,dg) ))$ the bounded operator valued functions over $\ca$. This forms a $C^*$-algebra with the norm
$$\| \Psi \| =  \sup_{A \in \ca} \{\|  \Psi (A )\| \}, \quad \Psi \in  \cf (\ca , \mathbb{B} (L^2(M, S,dg )) ). $$

\begin{definition}
Let
$$C =   \hbox{span} \{ e^X |   X \hbox{ vectorfield on  } M\}  . $$
The holonomy-diffeomorphism algebra $\mathbf{H D} (M,S,\ca)   $ is defined to be the $C^*$-subalgebra of  $\cf (\ca , \cb (L^2(M,S,dg )) )$ generated by $C$.
We will often denote $\mathbf{H D} (M,S,\ca)   $ by  $\mathbf{H D}  (M)$ when it is clear which $S$ and $\ca$ is meant.
\end{definition}

It was shown in \cite{AGnew} that  $\mathbf{H D} (M,S,\ca)   $ is independent of the metric $g$. \\

\subsection{The quantum holonomy-diffeomorphism algebra}

Let $\mathfrak{g}$ be the Lie-algebra of $G$.   
A section $\omega \in \Omega^1(M,\mathfrak{g})$ induces a transformation of $\ca$, and therefore an operator $U_\omega $ on $\mathcal{F}(\ca,  \mathbb{B} (L^2 (M ,  S,g)))$ via   
$$U_\omega (\xi )(A) = \xi (A + \omega) ,$$ 
which satisfy the relation 
\begin{equation} \label{konj}
(U_{\omega} e^X U_{ \omega}^{-1}) (A) = e^X (A + \omega )  .
\end{equation}
%
%
%
%
Finally note that infinitesimal translations on $\ca$ are formally given by 
\begin{equation}
E_\omega  =\frac{d}{dt}U_{  t  \omega}\Big|_{t=0} \;,
\label{soevnloes}
\end{equation}
where  
$$
E_{\omega_1+\omega_2}=E_{\omega_1}+E_{\omega_2\;,}
$$
which follows since the map $\Omega^1 (M,\mathfrak{g})\ni \omega \to U_{ \omega}$ is a group homomorphism, i.e. $U_{(\omega_1+\omega_2 )}=U_{\omega_1}U_{ \omega_2}$.

\begin{definition}
We define the $\mathbf{QHD}(M)$ as the algebra generated by elements in $\mathbf{HD}(M)$ and by all translations $U_{\omega}$, $\oo\in\OO^1(M,\mathfrak{g})$. 
\end{definition}

A priory $\mathbf{QHD}(M)$ is not a $*$-algebra. We will make $\mathbf{QHD}(M)$ into a $*$-algebra by requiring the $U_\omega$'s to be unitary. In particular will also require any representation to be a $*$-representation, in particular when we construct the Hilbert space the $U_\omega$'s has to be  unitary operators.


\subsection{Canonical commutation relations as operators on functions on $\ca$ }

In order to see why the quantum holonomy diffeomorphism algebra encodes the canonical commutation relations we will first discuss how these are realized on a space of functions on $\ca$. We will also need this discussion for later arguments. The following computations are purely formal.

To recap: The canonical quantization relations for a gauge theory are:
$$  [ \hat{E^\nu_b}({\bf m}_2) ,  \hat{A_\mu^a}({\bf m}_1) ] = i\hbar \delta_\mu^\nu \delta^a_b\delta ({\bf m}_1-{\bf m}_2)  , $$
where $(\hat{E},\hat{A})$ is the quantization of a pair of conjugate fields $(E,A)$.
If we have some Hilbert space $L^2 (\ca )$, then we would realize 
$\hat{A_\mu^a}({\bf m}_1)$ as
$$(\hat{A_\mu^a}({\bf m}_1)(\eta))(A) =A_\mu^a ({\bf m}_1) \eta (A), \quad \eta \in L^2 (\ca )$$
and we would realize $\hat{E^\nu_b}({\bf m}_2)$   as
\begin{eqnarray*}
( \hat{E^\nu_b}({\bf m}_2) \eta )(A)=\frac{d}{dt} \eta (A+t dm^\nu \sigma_b  \delta_{{\bf m}_2})|_{t=0}
\end{eqnarray*}
where $\delta_{{\bf m}_2}$ denotes the delta function localized in ${\bf m}_2$, 
and where $\sigma$ is a generator of $\mathfrak{g}$. Note that $A+t dx^\nu \sigma_b  \delta_{{\bf m}_2}$ is not a smooth connection, which will cause problems, if we were to realize $\hat{E}$ in a rigorous manner. Nevertheless, if these operators did exist they would realize the canonical commutation relations since
\begin{eqnarray*}
 \lefteqn{([ \hat{E^\nu_b}({\bf m}_2) ,  \hat{A_\mu^a}({\bf m}_1) ] \eta) (A)}\\
 &=&  \lim_{t \to 0}  \frac{(A +t d { m}^\nu \sigma_b \delta_{{ \bf m}_2}) _\mu^a({\bf m}_1)\eta (A +t d { m}^\nu \sigma_b \delta_{{ \bf m}_2})- A_\mu^a ({\bf m}_1)\eta (A )     }{t} \\
 && - A^a_\mu ({\bf m}_1) \lim_{t \to 0} \left( \frac{\eta (A +t d { m}^\nu \sigma_b \delta_{{ \bf m}_2})- \eta (A )     }{t}\right) \\
 &=&\delta_\mu^\nu \delta^a_b\delta ({\bf m}_1-{\bf m}_2)\eta (A) . 
 \end{eqnarray*}

This formal computation shows how singular the canonical commutation relations in fact are. In section \ref{conqft} we shall see that a modified version of the canonical commutation relations emerge from a more rigorous formalism that involves a representation of the $\mathbf{QHD}(M)$ algebra.

\subsection{The canonical commutation relations and the  $\mathbf{QHD}(M)$ algebra}
\label{seccan}

We will in this section see, how we can express $\hat{E^\nu_b}({\bf m}_2)$ and  $\hat{A_\mu^a}({\bf m}_1) $ with elements in the  $\mathbf{QHD}(M)$ algebra.

If we look at the operator $E_\omega$, then it is defined as $\frac{d}{dt} U_{t\omega} |_{t=0}$, and since $U_{t\omega} \xi (A)=\xi (A+\omega)$ it follows, that $E_\omega$ is essentially the same as  $\hat{E^\nu_b}({\bf m}_2)$, the only difference being that we have replaced $\omega^1\in \Omega (M,\mathfrak{g})$ with the singular one-form   $ d m^\nu \sigma_b  \delta_{{\bf m}_2}$. We can therefore express $\hat{E^\nu_b}({\bf m}_2)$ in terms of the $E_\omega$'s, simply by approximating  
 $ d m^\nu \sigma_b  \delta_{{\bf m}_2}$ with suitably elements in $\Omega^1 (M,\mathfrak{g})$. As an example, we can take a $L^2$ orthonormal basis  $\{ e_i\}$ for  $\Omega^1 (M,\mathfrak{g})$ and formally expand $ d m^\nu \sigma_b  \delta_{{\bf m}_2}$ in this basis, i.e.
 $$ d m^\nu \sigma_b  \delta_{{\bf m}_2} =\sum_{i}   (e_i)^\nu_b  ({\bf m_2 }) e_i , $$
 and in this way we get
 $$\hat{E^\nu_b}({\bf m}_2)= \sum_{i}   (e_i)^\nu_b  ({\bf m_2 }) E_{e_i} .  $$
Note that we can also go in the other direction: We can write
$$  \omega = \int_M (\omega ({\bf m}))_\nu^b dm^\nu \sigma_b \delta_{\bf m} d   {\bf m}      ,$$
and we thus get
$$ E_\omega = \int_M (\omega ({\bf m}))_\nu^b  \hat{E}_b^\nu ({ \bf m}) d   {\bf m}  .  $$

We now turn to $\hat{A}^a_\mu ({\bf m_1})$. First we consider $e^X$ for a given vectorfield $X$. These operators act on  functions in two variables, namely $A\in \ca$ and ${\bf m} \in M$.  We get
$$\frac{d}{dt} e^{tX} \eta (A,{\bf m})|_{t=0}=(X({\bf m})+A_{X({\bf m})})  \eta (A,{\bf m})  ,$$
where the notion $X({\bf m})$ is the derivative   $X({\bf m})$ in ${\bf m}$ acting in the second variable of $\eta$.  If we thus define the covariant derivative $\nabla^A=d+A$ we get 
$$\frac{d}{dt} e^{tX} \eta (A,{\bf m})|_{t=0}=\nabla^A_X  \eta (A,{\bf m})  .$$
Thus, as one would expect, an infinitesimal holonomy-diffeomorphism gives us a covariant derivative in the point ${\bf m}$. Although the operator $\hat{A}({\bf m})$ by itself is not an important physical entity we can of course extract it by considering a function $\eta$, which is constant in ${\bf m}$. In that case we get 
$$\frac{d}{dt} e^{tX} \eta (A,{\bf m})|_{t=0}= A_X ({\bf m})  \eta (A,{\bf m})  ,$$
or if we more specific chooses the vector field $X=\partial_\mu$ we get 
$$\frac{d}{dt} e^{tX} \eta (A,{\bf m})|_{t=0}= A_\mu  ({\bf m})  \eta (A,{\bf m})  .$$

Note, however, that the operator $\hat{A}^a_\mu ({\bf m})$ will always have a singular nature since evaluation at a point ${\bf m}$ in an $L^2$-space is not well defined (or rather: it has measure zero). This will also be reflected in the Hilbert space representation, that we construct in section \ref{sechilbert} (as well as in the Hilbert space representation that we constructed in \cite{Aastrup:2017vrm}).

To summarise, we find that the $\mathbf{QHD}(M)$ algebra naturally encodes the canonical commutation relations of a gauge theory in a manner that circumvents the otherwise singular nature of these relations.

\section{A gauge-invariant metric on $T\ca$}
\label{sektion_metrik_paa_a}

In this section we construct a gauge-invariant metric on the tangent bundle of $\ca$. This metric will play a key role in the subsequent sections, both in the construction of the Bott-Dirac operator in section \ref{secdirac} and in the construction of the Hilbert space in section \ref{sechilbert}.  \\

Let again $G$ be a compact Lie group, $\rho :G\to M_j (\mathbb{C}) $ a unitary faithful representation, $V=M\times \mathbb{C}^j$ a vector-bundle\footnote{We chose a complex bundle in order to be able to accommodate for instance $SU(n)$ and we chose it to be trivial in order to simplify the following analysis. Neither of these two conditions are believed to be important restrictions and can therefore be eased.} over the manifold $M$ and $\ca$ the space of $G$-connections acting in $V$. On the Lie algebra $\mathfrak{g}$ we have a real scalar product given by
$$(\rho (v),\rho (w))=Tr_j (vw^*) .$$
Note that this scalar product has the property
$$ (u\rho (v)u^*,u\rho (w)u^*)=Tr_j (vw^*)=(\rho (v),\rho (w))$$
for all unitaries $u\in M_j (\mathbb{C})$.

Our aim is to construct a metric on $T\ca$, the tangent space of $\ca$, and to do this we need to consider vector fields on $\ca$. Given an element $\xi\in \OO^1(M, \mathfrak{g})$, where $\mathfrak{g}$ is the Lie-algebra of $G$, we formally get a vector field on $\ca$ simply via 
\begin{equation}
\frac{\pa f}{\pa \xi} = \lim_{t\rightarrow 0} \frac{f(\nabla + t\xi) - f(\nabla)}{t}, \quad f\in C^\infty(\ca), \quad \nabla \in \ca . \label{tangent}
\end{equation}
In order to make this more stringent we need to specify what manifold structure we put on $\ca$.

We first consider $\Omega^1(M,\mathfrak{g})$. If we have an open subset $V\subset \mathbb{R}^n$ and a chart $\phi : V\to M$ we can define the semi-norm 
$$ \| \xi\|_{\phi, \alpha }:= \sup_{x\in U} ((\partial^\alpha \phi^{-1}(\xi))(x),(\partial^\alpha \phi^{-1}(\xi))(x))^{\frac{1}{2}},$$
where $\alpha \in \mathbb{N}_0^n$ and  $\xi\in \Omega^1(M,\mathfrak{g})$. If we have a finite atlas the system of semi-norms defined by the charts in the atlas gives $\Omega^1(M,\mathfrak{g})$ the structure of a Frechet space. The Frechet space structure is independent of the choice of a finite atlas. 

For a fixed connection $\nabla_0$ we define the map
$$\varphi_{\nabla_0} : \Omega^1(M,\mathfrak{g}) \to \ca$$
via 
$$\varphi_{\nabla_0} (\omega )=\nabla_0 +\omega.$$
Note that  $\varphi_{\nabla_0}$ is bijective.

\begin{proposition}
The system of maps $\{ \varphi_{\nabla_0}\}_{\nabla_0 \in \ca} $ equips $\ca$ with the structure of a smooth Frechet manifold. 
\end{proposition}

\textit{Proof.} We just need to prove that for two connections $\nabla_0,\nabla_1$ the map 
$$ \varphi_{\nabla_1}^{-1}\circ \varphi_{\nabla_0}:\Omega^1(M,\mathfrak{g})\to \Omega^1(M,\mathfrak{g})$$
is smooth. This is however clear, since it is given by translation with the one form $\nabla_0-\nabla_1 $. \\

Consequently we see that the map (\ref{tangent}) gives an isomorphism between $T_{\nabla_0} \ca$ and $\Omega^1(M,\mathfrak{g})$.    


Note that we have the adjoint action of $\frak{g}$ on $\frak{g}$, i.e.  $\frak{g}$ acts on $\frak{g}$ as commutators. In particular  a connection $\nabla \in \ca$ also acts on the bundle $M\times \frak{g}$. 
 
In the following we view a connection $\nabla \in \ca$ as a covariant derivative and thus it acts as 
\begin{equation}
\nabla : \Omega^* (M,\frak{g})\to \Omega^{*+1} (M,\frak{g}) . \label{covder}
\end{equation}
For readers more familiar with the local one form $A$ of a connection $\nabla$ we have $\nabla =d+A$.

The adjoint of $\nabla$ with respect to the scalar product on $\Omega^*(M,\mathfrak{g})$ given by
\begin{equation}
\int_M (\xi(m),\eta (m))dm  \label{inner1}
\end{equation}
where $(\cdot ,\cdot )$ is the scalar product induced by the chosen metric $g$ on $M$ and the scalar product on $\mathfrak{g}$,  acts as 
$$\nabla^* : \Omega^{*} (M,\frak{g})\to \Omega^{*-1} (M,\frak{g}) . $$
We therefore get the covariant Laplace operator
$$ \Delta_\nabla :=\nabla \nabla^* +\nabla^* \nabla :\Omega^{*} (M,\frak{g})\to \Omega^{*} (M,\frak{g}).$$
We here need both terms $\nabla \nabla^*$ and $\nabla^* \nabla$ to make the operator elliptic.  We will for construction of the metric consider $\Delta_\nabla$ restricted to $\Omega^{*} (M,\frak{g})$.


We define the scalar product on $T_\nabla \ca$ as follows
\begin{equation}
\langle \xi \vert \eta \rangle_\nabla = \langle (  1 + \tau_1\Delta_\nabla^p) \xi         , \left(  1 + \tau_1\Delta_\nabla^p\right) \eta   \rangle,
\label{inner}
\end{equation}
where $\xi,\eta\in\OO^1(M,\mathfrak{g})$, and where  $p$ and $\tau_1$ are fixed real positive parameters.

Note that the inner product (\ref{inner}) depends on $\nabla \in \ca$, i.e. for each $\nabla$ we have a different inner product.\\

Let us now consider a gauge transformation $G$ given by a unitary element $U\in C^\infty(M,G)$. When we consider a connection $\nabla$ as an operator in (\ref{covder}) it  transforms according to
$$
G(\nabla)= U^{-1}\nabla U,
$$
where the product is the operator product.

To see how ${\xi}\in T_\nabla\ca$ transforms under the gauge transformation we remind the reader that given a smooth map $\varphi : N_1 \to N_2$ between two manifolds $N_1$, $N_2$, a tangent vector $v$  in $T_nN_1$, $n\in N_1$, represented by a curve $\gamma$ with $\gamma '(0)=v$ transforms under the differential $D\varphi$ according to $D\varphi (v)=( \varphi\circ \gamma)'(0)$. At the level of derivations, $D\varphi (v)\in T_{\varphi (n)}N_2$ is given by the derivation 
$$ \lim_{t\rightarrow 0} \frac{f(\varphi (\gamma (t))-f(\varphi(\gamma (0)))}{t}.$$ 

Consequently 
\begin{eqnarray}
\frac{\pa f}{\pa DG(\xi)} \left( G(\nabla)  \right)
&=& \lim_{t\rightarrow 0} \frac{f(G(\nabla+t\xi)) - f(G(\nabla))}{t}
\nn\\
&=& \lim_{t\rightarrow 0} \frac{f(G(\nabla)+tU^{-1}\xi U)) - f(G(\nabla))}{t}
\nn\\
&=& \frac{\pa f}{\pa (U^{-1}\xi U)} \left( G(\nabla)  \right),
\nn
\end{eqnarray}
which implies that $DG({\xi})= U^{-1}\xi U$.

We can now compute how the inner product transforms under a gauge transformation: 
First we note that with respect to the inner product (\ref{inner1}) we have $U^{-1}=U^*$ and it thus follows $ (U^{-1}\nabla^* U )=(U^{-1}\nabla U )^*$.


A computation gives
\begin{eqnarray*}
DG(\Delta_\nabla\xi) & = & U^{-1}((\nabla\nabla^*+\nabla^*\nabla )\xi )U \\
&=&  U^{-1}(\nabla U U^{-1} \nabla^*U +\nabla^* UU^{-1}\nabla U)U^{-1} \xi )U \\
&=& (G(\nabla ) G(\nabla )^*+ G(\nabla )^* G(\nabla ) )DG(\xi) \\
&=& \Delta_{G(\nabla)} DG(\xi ) ,
\end{eqnarray*} 
and we thus finally get
\begin{eqnarray*}
\langle \xi , \eta \rangle_\nabla &=&\left\langle \left(  1 + \tau_1\Delta_\nabla^p \right) \xi         , \left(  1 + \tau_1\Delta_\nabla^p\right) \eta   \right\rangle \\
&=& \left\langle  U^{-1}\left( 1 + \tau_1\Delta_\nabla^p \right) \xi U         , U^{-1}\left(  1 + \tau_1\Delta_\nabla^p\right) \eta U   \right\rangle \\
&=&\left\langle DG \left( \left(  1 + \tau_1\Delta_\nabla^p \right) \xi \right)        , DG \left( \left(  1 + \tau_1\Delta_\nabla^p\right) \eta \right)   \right\rangle \\
&=&\left\langle \left(  1 + \tau_1\Delta_{G(\nabla)}^p \right) DG( \xi )        , \left(  1 + \tau_1\Delta_{G(\nabla)}^p\right) DG(\eta)    \right\rangle \\
&=& \langle DG(\xi),DG(\eta )\rangle_{G(\nabla)} .
\end{eqnarray*}
In particular the scalar product is gauge invariant.


\begin{remark}
The choice of the inner product (\ref{inner}) is essential. If we instead of the covariant Hodge-Laplace operator define  (\ref{inner}) with the ordinary Hodge-Laplace operator, then the result will be the Sobolev norm, which we used in \cite{Aastrup:2017vrm}. The role of the Hodge-Laplace operator in (\ref{inner}) is the same as in the Sobolev norm in \cite{Aastrup:2017vrm}, namely to serve as an ultra-violet dampening by suppressing modes below the scale $\tau_1$.   
\end{remark}

\begin{remark}
Note that (\ref{inner}) can be chosen in many different ways. In section \ref{conqft} we discuss the more general case where instead of $f(x)=(1+\tau_1 x^p)^{-1}$ we consider a function $f(x)$ with $\lim_{x\rightarrow\infty}f(x)=0$.
\end{remark}

\section{The Dirac and Bott-Dirac operators}
\label{secdirac}

Once we have the inner product (\ref{inner}) we can define the CAR bundle over $\ca$ by taking the complexified $\Lambda^* T\ca$. 
Denote by $\mbox{ext}(\xi)$ the operator of external multiplication with $\xi$ on $\Lambda^*T\ca$, where $\xi$ is an element in $T\ca$, and denote by $\mbox{int}(\xi)$ its adjoint, i.e. the interior multiplication by $\xi$.
Denote by $\{\xi_i\}$ a local orthonormal frame in $T\ca$ and let $\bar{c}_i$ and $c_i$ be the Clifford multiplication operators given by
\begin{eqnarray}
{c}_i &=& \mbox{ext}(\xi_i) + \mbox{int}(\xi_i)
\nn\\
\bar{c}_i &=& \mbox{ext}(\xi_i) - \mbox{int}(\xi_i) 
\end{eqnarray}
that satisfy the relations 
\begin{eqnarray}
 \{c_i, \bar{c}_j\} = 0, \quad
 \{c_i, {c_j}\} = \d_{ij}, \quad
 \{\bar{c}_i, \bar{c}_j\} =- \d_{ij}.
 \label{mangec}
\end{eqnarray}
%
%
%
We can now define the infinite-dimensional Dirac operator 
\begin{equation}
D f =\sum_i c_{i} \frac{\pa f}{\pa \xi_i}, \quad f\in C^\infty(\ca, \Lambda^* T\ca)
\label{dirac1}
\end{equation}
Since the gauge transformations act orthogonal on $T\ca$ this operator is invariant under gauge transformations in the sense that if we gauge transform the basis $\{ \xi_i\}$ with a gauge transformation $U$, i.e. if $\mu_j=\sum_iU_{ji}\xi_i$ we have, with $\tilde{c}_j= \mbox{ext}(\mu_j) + \mbox{int}(\mu_j) $:
 \begin{eqnarray*}
 D^U f &:=&\sum_j \tilde{c}_{j} \frac{\pa f}{\pa \mu_j}= \sum_j \sum_i \sum_k U_{ji} c_i  U^{-1}_{jk} \frac{\pa f}{\pa  \xi_k}  \\
 &=&  \sum_i \sum_k \delta_{ik} c_i   \frac{\pa f}{\pa  \xi_k} =\sum_i c_{i} \frac{\pa f}{\pa \xi_i} = Df.
 \end{eqnarray*}

The construction of the Dirac operator (\ref{dirac1}) raises a question concerning the expression $\frac{\pa f}{\pa \xi_i}$, which requires a trivialization of $T\ca$. Here we can either choose a global orthogonal frame, in which case (\ref{dirac1}) makes sense as it stands albeit it depends on the global frame. To see how it depends on the global frame let $\{\mu_j \}$ be a another frame. We can hence write 
$$\xi_i (\nabla)=\sum_{j}u_{ij}(\nabla)\mu_j (\nabla) ,$$ 
where $U(\nabla)=(u_{ij}(\nabla))$ is a family of orthogonal operators over $\ca$.  Thus given a section $\gamma$ in $T\ca$ we write
$$\gamma (\nabla) =\sum_i a_i(\nabla)\xi_i( \nabla)= \sum_{j,i} a_i(\nabla)u_{ij}(\nabla)\mu_j (\nabla).$$
Thus applying $D$ in the $\{\mu_j \}_j$ trivialization we get\footnote{Due to the computation above for the gauge invariance it is irrelevant if we use $\xi_i$ or $\mu_i$ for defining $D$}
$$ D\gamma =\sum_{k,i,j}c_{k} \left( \frac{\pa a_i}{\pa \xi_k} u_{ij}\mu_j +\frac{\pa u_{ij}}{\pa \xi_k} a_i\mu_j \right).$$
Thus in comparison to the $\{\xi_j \}_j$ trivialization we get the extra term 
$$\sum_{k,i,j}c_{k}  \frac{\pa u_{ij}}{\pa \xi_k} a_i\mu_j . $$
 Alternatively we can define the operator
\begin{equation}
D f =\sum_i c_{i} \nabla_{ \xi_i}f, \quad f\in C^\infty(\ca, \Lambda^* T\ca)
\label{dirac2}
\end{equation}
where $\nabla_{ \xi_i}$ denotes the Levi-Civita connection in $T\ca$. However the existence of the Levi-Civita connection in the infinite dimensional case is non-trivial and requires more computations.

\subsection{The Bott-Dirac operator}

We are now going to define an infinite-dimensional Bott-Dirac type operator. As with the Dirac operator we here have a choice between choosing a global orthogonal frame or working with a Levi-Civita connection. For simplicity we first choose the first option; we shall discuss the latter option in section \ref{levi}.

We define the Bott-Dirac operator by
\begin{equation}
B =\sum_i \left(c_{i} \tau_2\frac{\pa }{\pa \xi_i} + \bar{c}_i d_i  \right)
\label{BD}
\end{equation}
where $d_i\in C^\infty(\ca)$ is some function, which we will specify below, and where $\tau_2$ is a real parameter. 
The square of $B$ reads
\begin{equation}
B^2= \sum_{i}   \left(- \tau_2^2 \frac{\pa^2}{\pa \xi_i^2} + d_i^2 \right) + \tau_2 \sum_{ij} \bar{c}_i  c_j \left[ \frac{\pa}{\pa \xi_i} ,d_j \right].
\label{udsigt}
\end{equation}
Note that the function 
\begin{equation}
C^\infty(\ca)\ni\Psi(A) =  \exp\left( - \tau_2^{-1} S(A)  \right)
\label{groundstate}
\end{equation}
where $S(A)\in C^\infty(\ca)$, lies in the kernel of $B$, i.e.
\begin{equation}
B \Psi(A) =0,
\label{kernel}
\end{equation}
whenever
\begin{equation}
d_i = \frac{\pa S }{\pa \xi_i}.
\label{heat}
\end{equation}
The function $d_i$ will be specified in the following sections.
Note also that $B$ will be a gauge invariant operator whenever $S$ is gauge invariant. \\

\begin{remark}
The Bott-Dirac operator in (\ref{BD}) resembles the Bott-Dirac operator that Higson and Kasparov constructed in \cite{Higson}. To understand how the two operators are related we replace the covariant Hodge-Laplace operator $\Delta_A$ with the ordinary Hodge-Laplace operator $\Delta$ in (\ref{inner}). In this case ${\xi_i}$ becomes a global orthonormal frame in $T\ca$ and we can define a global coordinate system and expand $A$ according to $A=A_0+\sum_i x_i\xi_i$ (for details see \cite{Aastrup:2017atr}). In this case the linear part of $d_i$ will be of the form $s_i x_i$, and the Bott-Dirac operator $B$  will correspond to the one constructed by Higson and Kasparov in \cite{Higson}. Here the series $\{s_i\}$ corresponds to the momentum of a free field on a compact manifold. Note that in this case the square of $B$ is identical to the Hamiltonian operator of an infinite-dimensional harmonic oscillator, which we in \cite{Aastrup:2017atr} showed is identical to the Hamiltonian operator of the free sector of a gauge field coupled to a fermionic field.
\end{remark}

\section{The Hilbert space}
\label{sechilbert}

In this section we will  first discuss how to construct $L^2(\ca)$ in way, that involves a gauge fixing procedure, and secondly how to obtain a representation of the $\mathbf{HD}(M)$ algebra in a Hilbert space that also includes the Bott-Dirac operator.\\ 

For the construction of the Hilbert space we need two components:
\begin{enumerate}
    \item 
A gauge condition $\cf(A)=0$, which we assume that for some fixed connection $\nabla_0$ fulfills
$$\cf (\nabla_0+\lambda_1A_1+\lambda_2A_2)=\lambda_1 \cf (\nabla_0+A_1)+\lambda_2 \cf (\nabla_0+A_2) $$
Note that this just means that  we choose a connection $\nabla_0$ and identify $\ca $ with $\Omega^1 (M,\mathfrak{g})$
via
$$A=\nabla_0+A_1 \to A_1 ,$$
and that $A_1\to \cf ( \nabla_0+A_1)$ is linear. 
The example we have in mind is the following: $\cf (d+A)=\partial^\mu A_\mu =0$. Here $\nabla_0$ is the $0$-connection in the chosen trivialization.  
     \item
A positive gauge invariant function $S:\ca \to \bbR$. We will in section \ref{secS} discuss, how to choose $S$.    
\end{enumerate}

\subsection{Construction of $L^2(\ca )$}

We choose some $L^2$ basis $\{ \xi_i\}_{i\in \mathbb{N}}$ of $\Omega^1(M,\mathfrak{g})$, and write a connection
$$A=\nabla_0+\sum_{i=1}^\infty  x_i\xi_i  .$$ 
We denote by $\ca_n$ the space of connections of the form
$$A=\nabla_0+\sum_{i=1}^n x_i\xi_i.  $$
We identify $\ca_n$ with $\mathbb{R}^n$ via 
$$ \Phi (x_1,\ldots , x_n)=\nabla_0+x_1\xi_1+\ldots + x_n\xi_n .$$
This identification gives us a map $\mbox{P}_{n,n+1}: \ca_{n+1} \to \ca_n$ defined as 
$$\mbox{P}_{n,n+1} (x_1,\ldots ,x_n,x_{n+1}))=(x_1,\ldots ,x_n) .$$
Given a function $\eta$ on $\ca_n$, this gives rise to a function $\mbox{P}_{n,n+1}^* \eta$ on $\ca_{n+1}$ via 
$$( \mbox{P}_{n,n+1}^* \eta ) (A)=\eta (\mbox{P}_{n,n+1}(A)) .$$
We will also iterate these maps to get maps $\mbox{P}_{n,n+m}:\ca_n\to \ca_{n+m}$.

We put 
\begin{equation}
\cn_n= \int_{\mathbb{R}^n}  \exp(-\tau_2^{-1}S(A))dx_1\cdots dx_n  .  \label{normlization1}
\end{equation}

\begin{remark} \label{illdefined}
Due to the gauge symmetry of $S$ this integral will in general not be well defined in the continuum limit $n\rightarrow\infty$. In the following we will discuss the implementation of a gauge fixing procedure.
\end{remark}

For two measurable bounded functions $\zeta$ and $\eta$ on $\ca_n$ we first define 
\begin{equation}
\langle \zeta ,\eta \rangle_n =\frac{1}{\cn_n}\int_{\mathbb{R}^n} \overline{\zeta (A)}\eta (A)  \exp(-\tau_2^{-1}S(A))dx_1\cdots dx_n \label{normlization2} ,
\end{equation}
and then define
\begin{equation} 
\langle \zeta ,\eta \rangle  = \lim_{m\to \infty}  \langle \mbox{P}_{n,n+m}^*(\zeta) ,\mbox{P}_{n,n+m}^*(\eta ) \rangle_{n+m} \label{normlization3}  .
\end{equation}

\begin{remark} 
Note, that when the gauge group is non-Abelean, then in general 
$$\langle \zeta ,\eta \rangle_n \not=\langle \mbox{P}_{n,n+m}^*(\zeta) ,\mbox{P}_{n,n+m}^*(\eta )\rangle_{n+m}   .$$
This means that we cannot construct $L^2(\ca)$ as a direct limit of Hilbert spaces. 
Secondly: The convergence of the expression will, apart from the aforementined gauge fixing, of course depend on $S$. 

\end{remark}

We now want to implement the above integrals with a gauge condition $\cf (A)=0$. The most natural thing to try, is to  consider 
$$\left(\ca  /  \cf \right)_n  =    \{ A\in \ca_n \ |\  \cf (A) =0 \} ,$$
and then define
$$\cn_n= \int_{\mathbb{R}^n} \delta ( \left(\ca  /  \cf \right)_n  ) \exp(-\tau_2^{-1}S(A))dx_1\cdots dx_n  $$
and 
\begin{equation}
\langle \zeta ,\eta \rangle_n =\frac{1}{\cn_n}\int_{\mathbb{R}^n} \overline{\zeta (A)}\eta (A)\delta ( \left(\ca  /  \cf \right)_n  )   \exp(-\tau_2^{-1}S(A))dx_1\cdots dx_n ,
\label{normlization22}
\end{equation}
 where  $\delta ( \left(\ca  /  \cf \right)_n)$ is the $\delta$-function localized on the set 
$ \left(\ca  /  \cf \right)_n$.

This construction can however be problematic: If the basis $\{ \xi_i \}$ is poorly chosen it might be the case, that  
$$ \left(\ca  /  \cf \right)_n =\{ 0  \} .$$
This would leave the integrals meaningless. We therefore need to adopt strategies to choose $\{ \xi_i\}$ properly. An obvious strategy is the following: We consider the subspace 
$$\ca / \cf = \{ A\in \ca \ | \ \cf (A)=0 \} $$
and choose an orthonormal basis $\{\psi_i \}$ of $\ca / \cf$ and the pick an orthogonal  basis $\{ \phi_i \}$ of the orthogonal complement of $\ca / \cf$ in $\ca$. We can then proceed like in (\ref{normlization1}), (\ref{normlization2}) and (\ref{normlization3}), but where we only integrate over the degrees of freedom associated with $\{ \phi_i \}$.

We will now in a concrete example show how this could be implemented.
We first make a short detour into Hodge theory: 
The exterior derivative 
$$d:\Omega^* (M)\to \Omega^{*+1} (M)$$ 
has an adjoint 
$$d^* :\Omega^{*+1} (M) \to \Omega^* (M).$$ 
Hodge theory states that we have an $L^2$-orthogonal decomposition
$$\Omega^*(M)= d (\Omega^{*-1}(M))\oplus  d^* (\Omega^{*+1}(M)) \oplus H^*(M) ,$$
where 
$$ H^*(M)=\hbox{Ker} \{ \Delta :\Omega^* (M) \to \Omega^* (M) \}, $$ 
where $\Delta =dd^* +d^* d$. Note that 
$$ H^*(M)=\hbox{Ker}\ d \cap \hbox{Ker}\ d^* , \quad \hbox{Ker}\ d =d (\Omega^{*-1}(M))\oplus   H^*(M).$$

This entire construction can be tensored with $\mathfrak{g}$. We then obtain an $L^2$-decomposition of $\ca \simeq \Omega^1(M, \mathfrak{g})$ as
$$\Omega^1(M,\mathfrak{g})= d (\Omega^{0}(M,\mathfrak{g}))\oplus  d^* (\Omega^{2}(M,\mathfrak{g})) \oplus H^1(M,\mathfrak{g}) ,$$

As our gauge fixed space we take 
$$\ca / \cf = d^* (\Omega^{2}(M,\mathfrak{g}))  .$$
Since 
$$\hbox{ker}\ d^*= d^* (\Omega^2 (M,\mathfrak{g})))\oplus H^1(M,\mathfrak{g}) ,$$
modulo the cohomology $H^1(M,\mathfrak{g})$, this corresponds to the gauge condition  $d^* A =0$, i.e. $\cf(A) =d^* A $.  The exterior derivative is in local coordinates given via
$$d=\sum_{\mu} \hbox{ext}(dm_\mu )\frac{\partial }{\partial m_\mu} .$$
If we assume that we are in the flat case where the $x_i$ are metric coordinates, we get 
$$d^*= \sum_\mu -\hbox{int}(dm_\mu )\frac{\partial }{\partial m_\mu}  ,$$ since the adjoint of $\hbox{ext}(dm_\mu )$ is $\hbox{int}(dm_\mu )$.
Thus the gauge condition is 
$$\partial^\mu A_\mu =0 $$
with the extra condition, that we exclude the cohomology. From a BRST-viewpoint it is, however, quite natural to exclude the cohomology: If we for example consider a constant connection $\sigma_i dm^\mu$, then this connection is flat. This means that the intersection between the flat gauge orbit and solution to the equation  $\partial^\mu A_\mu =0 $  consists of more than one point. 

We can thus construct the bases $\{ \psi_i\}$ and $\{ \phi_i \}$ for the inner product in $L^2(\ca)$ by constructing eigenvectors for $\Delta$, and split them in the spaces $d^* (\Omega^{2}(M,\mathfrak{g}))$ and $d (\Omega^{0}(M,\mathfrak{g}))   \oplus H^1(M,\mathfrak{g}) .$

\subsection{Possible choices for $S$ }
\label{secS} 
 
We will now discuss the choice of $S$. The first choice we discuss is
\begin{equation} \label{f2}
S(A)=\int_M \hbox{Tr}( F(A)\star F(A) ),
\end{equation}
where $F(A)$ is the curvature, i.e.
$$F(A)=dA +A\wedge A ,$$
and $\star$ is the Hodge-star. 
If we look at the case of a  free theory we get 
$$S(A)=\int_M \hbox{Tr}( dA\star dA )=\langle dA ,dA \rangle =\langle A ,  d^* dA \rangle ,$$
where $\langle \cdot , \cdot \rangle $ denotes the inner product on $\Omega^* (M,\mathfrak{g})$ induced by the chosen metric on $M$. Upon implementing a gauge fixing as described in the previous section, this is
$$S(A)= \langle A ,  \Delta A \rangle,$$
where $\Delta$ is the Hodge-Laplace operator. This is very similar to the construction in \cite{Aastrup:2017atr} and \cite{Aastrup:2017vrm}, where\footnote{Note that the $p$ in formula (\ref{inner}) need not be the same as the $p$ in $S(A)$. If the two are different one will have to adjust the assessment of when a representation of the $\mathbf{QHD}(M)$ algebra exists accordingly.} 
\begin{equation} \label{gammel}
S(A)= \langle A+\Delta^p A ,A+  \Delta^p A \rangle 
\end{equation}
were used. In \cite{Aastrup:2017atr} it was proved that the representation of the $\mathbf{QHD}(M)$ algebra existed in dimension 3 when $p>\frac54$. The current case of $S(A)= \langle A ,  \Delta A \rangle$ would  correspond to $p=\frac12$, and we are therefore certain, that with this choice of $S$ the representation of the $\mathbf{QHD}(M)$ algebra will not converge. Since we expect the convergence issues to be worse in the case of an interacting theory, the choice (\ref{f2})
is not suitably. Instead it would be desirable to have something like (\ref{gammel}) where higher order derivatives are appearing, but in a covariant fashion. One possibility is the following:
Since $*F(A) \in \Omega^1 (M,\mathfrak{g})$ we can use $\Delta_A$ to construct
\begin{equation}
S(A) =\langle   \Delta_A^p ( \star F(A)) , \Delta_A^p (  \star F(A))   \rangle , \quad p \in ]0,\infty[,
\label{solution2}
\end{equation}
which is a covariant version of (\ref{gammel}).

Another possibility for choosing $S$, which is related to our present construction, is to use the metric on $\ca$ described in section \ref{sektion_metrik_paa_a}. In this case a natural candidate for $S$ is
\begin{equation}
S(A)=\inf \{ (d^2_{\ca} (A,B)) \ | \  B\in \ca , \ B \ \hbox{flat}  \}  .
\label{solution3}
\end{equation}
Here $d_{\ca} (A,B)$ denotes the distance from $A$ to $B$ induced by the metric on $\ca$. Since the metric is gauge covariant, the distance becomes gauge invariant. So $S(A)$ in (\ref{solution3}) is just the distance to the power $p$  from $A$ to the gauge orbit of flat connections.  

If we do this construction with the metric on $\ca $, but use the following scalar product (which is of course not covariant)  
\begin{equation}
\langle \xi \vert \eta \rangle_A = \int_M \left(  \left(  1 + \tau_1\Delta^p\right) \xi({\bf m})         , \left(  1 + \tau_1\Delta^p\right) \xi({\bf m})   \right)       dm,
\label{innerikke}
\end{equation}
instead of (\ref{inner}), then the scalar product is independent of $A$, and consequently the induced distance on $\ca$ is 
\begin{eqnarray*}
d(A,B)^2  & =& \langle A-B \vert A- B \rangle\\
&= &\int_M \left(  \left(  1 + \tau_1\Delta^p\right) (A-B)({\bf m})         , \left(  1 + \tau_1\Delta^p\right) (A-B)({\bf m})   \right)       dm,
\end{eqnarray*}
which is just (\ref{gammel}). We therefore expect (\ref{solution2}) and (\ref{solution3}) to be quite similar. Note they both involve an ultra-violet regulation in the form of a Hodge-Laplace operator to the power $p$. We believe, however, that the latter is the more canonical choice.

Combining the above lead us to the conjecture
\begin{conj}
\label{rach}
Let $S(A)$ be given by either equation (\ref{solution2}) or (\ref{solution3}). It is then possible in either case to choose $p$ sufficiently large so that the inner product (\ref{normlization22}) and the expectation values of the $\mathbf{HD}(M)$ algebra exist. 
\end{conj}

\begin{remark}
Generally speaking, both options (\ref{solution2}) and (\ref{solution3}) involve an ultra-violet regulation in the form of a covariant Hodge-Laplace operator to the power $p$. We know that the representation of the $\mathbf{QHD}(M)$ algebra exist in the non-covariant case  \cite{Aastrup:2017vrm} for $p>\frac{5}{4}$. Although we expect the convergence to be worse in the covariant case, we believe that conjecture \ref{rach} must be true.
\end{remark}

\subsection{Representation of the $\mathbf{HD}(M)$ algebra in the CAR bundle}

The aim is to represent the  $\mathbf{HD}(M)$ algebra in the Hilbert space 
$$
\ch= L^2(\ca)\otimes \Lambda^* T\ca.
$$
Note that since we have a metric on $ T\ca$ we can equip $\ch$ with a Hilbert space structure.

If we consider the space 
$$L^2 (\ca ) \otimes L^2 (M,S, dg) $$
we can represent the holonomy-diffeomorphism algebra using (\ref{chopin1}). There is of course the convergence issue of the holonomies as discussed the previous section. In order to represent $\mathbf{HD}(M)$ on  
$\ch= L^2(\ca)\otimes \Lambda^* T\ca
$ we first need, for a given connection $A$, to represent the $e^X_A$-operators on  $\OO^1(M,\mathfrak{g})$.  We therefore begin with the basespace 
\begin{equation}
\ch_b = \OO^1(M,\mathfrak{g})
\label{trmpp}
\end{equation}
where the Hilbert space structure is with respect to the norm (\ref{inner})
in the sense that the right-hand side of (\ref{trmpp}) has been completed in this norm. Note that this Hilbert space structure is connection dependent.

Let $\oo\in\OO^1(M,\mathfrak{g})$ and let $e^X$ be the flow of a vector field $X$ in $M$. Let ${\bf m}_1\in M$ and ${\bf m}_2= e^X({\bf m}_1)$, and $\gamma$ the path $t\rightarrow e^{tX}({\bf m}_1)$. Furthermore we denote by $(e^{-X})^*(\oo)$ the pullback of the one-form part of $\oo$ by the diffeomorphism $e^{-X}$, i.e. $(e^{-X})^*$ leaves the Lie algebra $\mathfrak{g}$ unchanged\footnote{Here another possibility is to use the Levi-Civita connection of the metric $g$ to parallel transport the one-form part of $\oo$. If this is done the representation of the $\mathbf{HD}(M)$ algebra will still be non-unitary (see the discussion below) but the non-unitarity will be of the order of $\tau_1^{1/2p}$, i.e. the $L^2$ part of the representation will be unitary.}. We define
$$
e^X_A(\oo)({\bf m}_2)= \mbox{Hol}(\gamma,A) \left( (e^{-X})^*(\oo)({\bf m}_2)  \right) (\mbox{Hol}(\gamma,A))^{-1}
$$
which we extend to an action on elements in $\Lambda^* \OO^1(M,\mathfrak{g})$ by
$$
e^X_A(v_1\wedge \ldots\wedge v_n) = e^X_A(v_1)\wedge \ldots\wedge e^X_A(v_n) .
$$
The action on $\eta\in L^2 (\ca)\otimes \Lambda^* T_A\ca$ is then defined by
\begin{equation}
e^X(\xi\otimes\oo)(A)= \xi(A)e^X_A(\oo),
\label{actbundle}
\end{equation}
with $  \eta = \xi\otimes\oo$. This gives us an action of the $\mathbf{HD}(M)$ algebra on the CAR-bundle\footnote{Technically speaking we do not get a representation of the $\mathbf{HD}(M)$ algebra, since this  algebra is defined using unitary representations. Rather we get a representation of a non unitary version of the flow algebra, see \cite{AGnew} for details. }, which we straight forwardly extend to the full $\mathbf{QHD}(M)$ algebra via
$$
U_\oo(\xi\otimes\oo)(A) = (\xi\otimes\oo)(A-\oo).
$$
These definitions were discussed in \cite{Aastrup:2018coc}. Note that this in general not a unitary action of  $\mathbf{HD}(M)$  on  $L^2 (\ca)\otimes \Lambda^* T_A\ca$. The first problem is that the flow $X$ might not preserve the metric on $M$, and hence the flow on does not act isometric on $\Omega^1(M)$. On top of this comes that we are working with a norm induced by the covariant Hodge-Laplace operator and this will distort the unitarity  further.  When the action is not unitary on  $L^2 (\ca)\otimes  T_A\ca$, it will be not be bounded on  $\mathbf{HD}(M)$  on  $L^2 (\ca)\otimes \Lambda^* T_A\ca$ unless we restrict it to elements with finite particle number.

Next, we would like the action of $\mathbf{HD}(M)$ to be gauge invariant. This is a priory not the case, since sections in the CAR bundle need not be gauge covariant, whereas the holonomies are. We therefore need to introduce the notion of gauge-covariant section, which is a section $\eta$ that satisfies
$$
\eta(G(A)) = G(\eta)(A),
$$
i.e. on a gauge orbit the section $\eta$ takes values, which are compatible with the gauge transformation $G$. Thus, when the $\mathbf{HD}(M)$ algebra is represented on gauge-covariant sections the representation is gauge-covariant. Note, however, that this does not apply to the full $\mathbf{QHD}(M)$ algebra since the translation operators $U_\oo$ do not preserve gauge-covariance. The Bott-Dirac operator, on the other hand, does.

Thus, to summarize we have a representation of the $\mathbf{HD}(M)$ algebra and the Bott-Dirac operator on the Hilbert space $\ch=L^2(\ca)\otimes \Lambda^*T\ca$, which in the case of the $\mathbf{HD}(M)$ algebra and the Bott-Dirac operator is gauge covariant.

\subsection{The BRST symmetry and the Bott-Dirac operator}

Let us briefly discuss an alternative approach to defining the Hilbert space and in particular the integral (\ref{normlization22}), that includes a BRST gauge fixing procedure \cite{Becchi:1975nq,Tyutin:1975qk,Piguet}. The aim is to make the construction of $L^2(\ca)$ independent of the choice of the gauge condition $\cf(A)$.

Let $(c,\bar{c})$ be a pair of Fadeev-Popov ghost and anti-ghost fields 
and denote by $b$ a Lautrup-Nakanishi field. The nilpotent BRST transformation reads 
\begin{equation}
\begin{array}{cc}
s A = D c &\quad s c = c^2
\nn\\
s \bar{c} = b &\quad sb =0.
\end{array}
\end{equation}

The improved version of the integral (\ref{normlization22}) is then written
\begin{equation}
\langle \zeta ,\eta \rangle_n =\frac{1}{\cn_n}\int  \cd \ca_n \cd c_n \cd\bar{c}_n \cd b_n \overline{\zeta (A)}\eta (A)  \exp(-S(A) + s\OO ) ,
\label{Lautrup}
\end{equation}
where the normalisation $\cn_n$ has also been modified in the same manner. Here $\cd \ca_n=dx_1\ldots dx_n$ and $ \cd c_n$, $\cd\bar{c}_n $ and $\cd b_n $ are the corresponding integration measures over the various fields truncated to the finite-dimensional case.
The additional exponent $\OO$ in (\ref{Lautrup}) is then of the form
$$
\OO = \int \bar{c}  (\cf(A)+ i b/2 )
$$
where $\cf(A)$ is again the gauge fixing condition.

Now, the point that we wish to discuss here is that the ghost sector will also show up in the Bott-Dirac operator (\ref{BD}) via the factor $\sum_i \bar{c}_i d_i$. Namely, in order for the ground state to lie in the kernel of $B$ the factor $d_i$ must now be set to
$$
d_i=\frac{\pa (S+s\OO)}{\pa\xi_i}.
$$
Ideally this will then be of the form 
$$
d_i = d_i\vert_{\mbox{\tiny physical}} + s k_i\vert_{\mbox{\tiny unphysical}} 
$$
which would be the case if the BRST transformation commutes with $\frac{\pa}{\pa \xi_i}$. This is, however, not the case since
$$
\left[s, \frac{\pa}{\pa \xi_i}\right] A = [c, \xi_i]  .
$$
If we instead define the nilpotent operator $\tilde{s}= s+\delta_{c}$, where 
$$
\delta_{c}\xi_i = [ \xi_i,c] 
$$
is simply the odd rotation of the basis $\{\xi_i\}$, then 
$$
\left[\tilde{s}, \frac{\pa}{\pa \xi_i}\right] =0.
$$
This means that $d_i$ will have the desire form
$$
d_i = \frac{\pa S}{\pa \xi_i} + \tilde{s} \frac{\pa \OO}{\pa \xi_i}.
$$
Thus, we end up with a Bott-Dirac operator, which involves a physical gauge-invariant sector and an unphysical ghost sector; the former being identified as the cohomology of  the nilpotent BRST operator $\tilde{s}$.

\section{A complex phase and the ground state}
\label{secCS}

The square of the Bott-Dirac operator in equation (\ref{udsigt}) involves the term $\sum_i d_i^2$ where $d_i  $ is given by the derivative of $S(A)$ (see equation (\ref{heat})). As we shall further discuss in section \ref{conqft}, $d_i$ will not involve terms linear in the curvature $F$ when $S(A)$ is defined as discussed in section \ref{secS}. For reasons that shall soon become clear we would however like $d_i$ to involve such linear terms. In this section we shall discuss one possible strategy to achieve this by introducing a complex phase.\\

Let us begin by introducing the Chern-Simons functional
$$
CS(A) =  \int_M \mbox{Tr} \left( A\wedge dA + \frac{2}{3} A\wedge A \wedge A  \right)
$$
 and noting that if we add a Chern-Simons term to $S(A)$ as a complex phase
\begin{equation}
S'(A) =S(A) + \frac{\mathrm{i}}{2} CS(A)
\label{first}
\end{equation}
then we get
$$
d_i = \frac{\mathrm{i}}{2} \frac{\pa  CS}{\pa \xi_i} + \ldots
$$
which, combined with
\begin{equation}
\frac{\pa CS}{\pa \xi_i} =  2 \int_M \mbox{Tr}\left(\xi_i\wedge F(A)  \right)
\label{godhavn}
\end{equation}
gives us a term in $d_i$ that is linear in $F(A)$. The addition of a complex phase to the function 
\begin{equation}
\Psi'(A) =  \exp\left( - \tau_2^{-1} S'(A)  \right)
\label{groundstate2}
\end{equation}
in (\ref{groundstate}) raises, however, the problem that the Bott-Dirac operator $B$ will now be a complex operator and that $B$ and $B^*$ will not have the same kernel. With definition (\ref{first}) the function $\Psi'(A)$ will lie in the kernel of $B$ but not in the kernel of $B^*$.

The solution to this problem is to double the Hilbert space 
$$
\ch= \ch^{(+)}\oplus \ch^{(-)},\quad \ch^{(\pm)} = L^2(\ca) \otimes \Lambda^* T\ca
$$ 
and write the function (\ref{groundstate2}) as the sum $\Phi=\Psi^{(+)}\oplus \Psi^{(-)}$ 
with
\begin{equation}
\Psi^{(+)}(A)  = \exp\left(- \tau_2^{-1} S^{(+)}(A)\right), \quad \Psi^{(-)}(A)= \exp\left(- \tau_2^{-1} S^{(-)}(A)\right)
\label{splus}
\end{equation}
with
\begin{equation}
S^{(\pm)}(A) = S(A) \pm \frac{i}{2}CS(A).
\end{equation}
Furthermore, we write 
\begin{equation}
B^{(\pm)} =\sum_i \left(c_{i} {\tau}_2\frac{\pa }{\pa \xi_i} + \bar{c}_i d^{(\pm)}_i  \right), \quad d_i^{(\pm)} = \frac{\pa S^{(\pm)}}{\pa\xi_i}
\label{hmm}
\end{equation}
that satisfy
\begin{equation}
\left( B^{(+)} \right)^* = B^{(-)},\quad  \left( B^{(-)}\right)^* = B^{(+)}
\end{equation}
as well as
\begin{equation}
B^{(+)}\Psi^{(+)}=0    ,\quad    B^{(-)}\Psi^{(-)}=0 .
\end{equation}
If we write the function $\Phi$ as
\begin{equation}
\Phi=
\left(
\begin{array}{c}
\Psi^{(+)}
\\
\Psi^{(-)}
\end{array}
\right)
\label{GS}
\end{equation}
and define the Bott-Dirac operator acting in $\ch$ with
$$
\cb =
\left(
\begin{array}{cc}
 0 & B^{(-)}
\\
 B^{(+)}  &  0
\end{array}
\right) ,
$$
then we obtain the desired relations
$$
\cb\Phi(A)=0,\quad \cb^*=\cb,
$$
i.e. the function $\Phi$ lies in the kernel of both $\cb$ and its adjoint.

Now, this construction does not quite do the job as the function $S(A)$ is used to define the Hilbert space inner product (\ref{normlization22}). This means that we cannot simply add a complex phase to $S(A)$ since this would result in an inner product that is not positive definite. What we can do instead is to change the inner product (\ref{normlization22}) by removing the factor $\exp(-\tau_2^{-1}S(A))$ and instead inserting it into the states. Thus, what we suggest is to map the states according to
$$
\eta(A) \rightarrow \eta(A)\exp(-\tau_2^{-1}S(A)).
$$
and then evaluate them with an inner product that does not include the factor $\exp(-\tau_2^{-1}S(A))$.
In this way the function (\ref{GS}) becomes the ground state, i.e. it satisfies the relation
$$
\left\langle \Phi, \cb \Phi\right\rangle_\ch =0
$$
where $\langle \cdot, \cdot \rangle_\ch$ is the inner product in $\ch$.

\begin{remark}
Let us point out the similarity between $\Psi_{\mbox{\tiny gs}}(A)$ and the Kodama state in canonical quantum gravity \cite{Kodama:1988yf} (see also \cite{Smolin:2002sz} and references therein). If we choose $\ca$ as a configuration space of Ashtekar connections \cite{Ashtekar:1986yd,Ashtekar:1987gu}, then $\Psi_{\mbox{\tiny gs}}(A)$ would strongly resemble the Kodama state, albeit the present setup is different from the one presented in \cite{Smolin:2002sz}.
\end{remark}

Next let us compute the square of $\cb$ 
$$
\cb^2 = 
\left(
\begin{array}{cc}
  B^{(-)}B^{(+)}   &0
\\
0& B^{(+)}   B^{(-)}
\end{array}
\right) 
$$
where
\begin{eqnarray}
 B^{(-)}B^{(+)} =   \sum_{i}   \left(- {\tau}_2^2 \frac{\pa^2}{\pa \xi_i^2} +  d_i^{(+)} d_i^{(-)} \right)+{\tau}_2 \sum_{ij} c_i\bar{c}_j \left( \frac{\pa}{\pa \xi_i}    d^{(+)}_j    -   d^{(-)}_j  \frac{\pa}{\pa \xi_i}  \right)    
\nn\\
B^{(+)}   B^{(-)} =  \sum_{i}   \left(- {\tau}_2^2 \frac{\pa^2}{\pa \xi_i^2} +  d_i^{(+)} d_i^{(-)} \right)+ {\tau}_2 \sum_{ij} c_i\bar{c}_j \left( \frac{\pa}{\pa \xi_i}    d^{(-)}_j    -   d^{(+)}_j  \frac{\pa}{\pa \xi_i}  \right)    
\end{eqnarray}
which we also write as
\begin{eqnarray}
 B^{(-)}B^{(+)} =  H_b+ H_f^{(+)}  
\nn\\
B^{(+)}   B^{(-)} = H_b + H_f^{(-)}  
\end{eqnarray}
where
\begin{equation}
H_b= \sum_{i}   \left(- {\tau}_2^2 \frac{\pa^2}{\pa \xi_i^2} +  d_i^{(+)} d_i^{(-)} \right),\quad H_f^{(\pm)} =  {\tau}_2 \sum_{ij} c_i\bar{c}_j \left( \frac{\pa}{\pa \xi_i}    d^{(\pm)}_j    -   d^{(\mp)}_j  \frac{\pa}{\pa \xi_i}  \right)   .
\label{Hb}
\end{equation}
In the next section we shall see that $H_b$ is the Hamiltonian of a quantised gauge theory.
Let us take a closer look at $H_f^{(\pm)}$. With
$$
d_j^{(\pm)} = \frac{\pa S}{\pa\xi_j} \pm  \frac{\mathrm{i}}{2}\frac{\pa CS}{\pa\xi_j}
$$
we obtain
\begin{eqnarray}
H_f^{(\pm)} &=&  {\tau}_2 \sum_{ij} c_i\bar{c}_j \left( \frac{\pa^2S}{\pa \xi_i\pa\xi_j}       + \frac{\mathrm{i}}{2}  \left\{  \frac{\pa}{\pa \xi_i} , \frac{\pa CS}{\pa \xi_j}   \right\}  \right)   
\nn\\
&=&  {\tau}_2 \sum_{ij} c_i\bar{c}_j \left( \frac{\pa^2 S}{\pa \xi_i\pa\xi_j}       + \frac{\mathrm{i}}{2}  \frac{\pa^2 CS}{\pa \xi_i\pa \xi_j}  +\mathrm{i} \frac{\pa CS}{\pa \xi_j}  \frac{\pa}{\pa \xi_i}     \right) ,
\label{Hf}
\end{eqnarray}
which we in the next section will tentatively identify as the Hamiltonian of a fermionic sector in a quantum field theory.

Finally note that there is a natural real structure
$$
J=\left(
\begin{array}{cc}
0 & C
\\
 C  &  0
\end{array}
\right) ,\quad J^2=\mathbf{1}_2,
$$
where $C$ denotes complex conjugation and where $\mathbf{1}_2 $ is the $2\times 2$ identity matrix, which gives us the relation 
$$
J\cb= \cb J.
$$

\section{The emergence of local quantum field theory}
\label{conqft}

In this section we will discuss how the representation of the $\mathbf{QHD}(M)$ algebra together with the Bott-Dirac operator in fact amounts to a quantum field theory of a gauge field coupled to a fermionic field on a curved background. \\


Let us first consider the limit $\tau_1=0$.
In section \ref{seccan} we demonstrated that the formal field operators $(\hat{A}({\bf m}),\hat{E}({\bf m}))$ satisfy the canonical commutation relations of a gauge theory. The operator $\hat{E}$ can be written
\begin{equation}
    \hat{E} ({\bf m}) =\sum_j  \xi_j ({\bf m}) \frac{\partial }{\partial \xi_j}  \label{vladp}
\end{equation}
with the reverse
$$\frac{\partial}{\partial \xi_j} = \int \xi_j({\bf m})\hat{E}({\bf m}) dm . $$
If we therefore consider $\sum_j \frac{\partial}{\partial \xi_j} \frac{\partial}{\partial \xi_j}$ we get
\begin{equation} 
\sum_j \frac{\partial}{\partial \xi_j} \frac{\partial}{\partial \xi_j} = \int \hat{E}({\bf m}) \hat{E}({\bf m}) dm , \quad (\tau_1=0)
\label{philipglass}
\end{equation}
where we used $\sum_j \xi_j({\bf m}_1)\xi_j({\bf m}_2)=\delta ({\bf m}_1-{\bf m}_2)$.

Let us now consider what happens when $\tau_1\not=0$. In this case the set $\{ \xi_i\}$ is orthonormal with respect to the Sobolev norm (\ref{inner}) on $\OO^1(M,\mathfrak{g})$, which is constructed via the covariant Laplace operator $\Delta_A$ for a given connection $A\in\ca$. 
Since the eigen-vectors $\xi_i$ now depend on $A$ we adopt the notation
$$
\hat{E}_A({\bf m})=\sum_j \xi_j({\bf m}) \frac{\partial }{\partial \xi_j} .
$$
Note, due to formula (\ref{vladp}), that $\hat{E}({\bf m})$ and $\hat{E}_A({\bf m})$ coincide in the limit $\tau_1 \to 0$. 

Let $\lambda_i$ be the eigenvalues of $\Delta_A$ and $\{e_i\}$ the associated $L^2$-eigenvectors. In the case of the norm (\ref{inner}) we have 
$$
\xi_i=\frac{e_i}{1+\tau_1\lambda_i^p}
$$
but we can in fact take any bounded  function $f:[0, \infty) \to \R$ with 
$$
\lim_{x\to 0} f(x)=0,
$$ 
instead of $f(x)=(1+\tau_1x^p)^{-1}$. Then
$$ K_{f,A}({\bf m}_1,{\bf m}_2)= \sum_j \xi_j({\bf m}_1)\xi_j({\bf m}_2)$$
is the integral kernel of $f^2(\Delta_A)$, i.e. 
$$f^2(\Delta_A)(\eta)({\bf m}_1)=\int_M K_{f,A} ({\bf m}_1,{\bf m}_2)\eta({\bf m}_2) dm_2 ,$$
which is easily seen by evaluating both sides on the $e_i$'s.

Let us now consider what happens to equation (\ref{philipglass}) when $\tau_1\not=0$. In that case we find the expression
\begin{eqnarray*}
\sum_j \frac{\partial}{\partial \xi_j} \frac{\partial}{\partial \xi_j} 
 & = &\int \int  \hat{E}({\bf m}_1) K_{f,A}({\bf m}_1,{\bf m}_2) \hat{E}({\bf m}_2)dm_1 dm_2 \\
 &=& \int (\hat{E}({\bf m}), f^2 (\Delta_A)(\hat{E})({\bf m})) d m
\end{eqnarray*}
Note here that we have the $\hat{E}$ operators as defined in section 2 and not the $\hat{E}_A$ operators. If we consider the limit $\tau_1\to 0$ there is an expansion in terms of kernels
$$  K_{f,A}({\bf m}_1,{\bf m}_2) \sim k_0({\bf m}_1,{\bf m}_2)+\tau_1^{{1/2p}} k_1({\bf m}_1,{\bf m}_2)+\ldots $$
where it is understood that $k_i$ depend on $A$ except for $k_0$, which is 
$$
k_0({\bf m}_1,{\bf m}_1)=\delta({\bf m}_1 - {\bf m}_2)\mathbf{1}_{\mathfrak{g}},
$$ 
where $\mathbf{1}_\mathfrak{g}$
is the identity in the Lie algebra.
Hence we get 
\begin{eqnarray}
\int \int  \hat{E}({\bf m}_1) K_{f,A}({\bf m}_1,{\bf m}_2) \hat{E}({\bf m}_2)dm_1 dm_2 
\hspace{-5cm}&&\nn\\
&\sim& \int \hat{E} ({\bf m}_1) k_0({\bf m}_1,{\bf m}_2)\hat{E} ({\bf m}_2)dm_1dm_2
\nn\\&&
+\tau_1^{{1/2p}} \int \hat{E} ({\bf m}_1) k_1 ({\bf m}_1,{\bf m}_2)\hat{E} ({\bf m}_2) dm_1 dm_2+\ldots   \nn \\
&=&         \int (\hat{E} ({\bf m}) ,\hat{E} ({\bf m}))dm \nn
\\
&& +\tau_1^{{1/2p}} \int (\hat{E} ({\bf m}), 2f'(0)f(0)\Delta_A( \hat{E}) ({\bf m})) dm +\ldots    
\label{Boris}
\end{eqnarray}

The integral kernel also turns up in the second term in $H_b$ in (\ref{Hb}). Let us here only consider the part that arises from the Chern-Simons term in (\ref{splus}), i.e.
$$
\sum_i d_i^{(+)} d^{(-)}_i = \sum_i \frac{1}{4}\left(\frac{\pa CS(A)}{\pa \xi_i}\right)^2+ \mbox{higher derivative terms}.
$$
where 'higher derivative terms' refer to those terms coming from $S(A)$.
Using (\ref{godhavn})
we obtain
\begin{eqnarray*}
\sum_i d_i^{(+)} d^{(-)}_i &=&  \int \int  K^{ab}_{f,A}({\bf m}_1,{\bf m}_2) F^a({\bf m}_1) F^b({\bf m}_2)  dm_1 dm_2
\nn\\
&&\hspace{2cm} + \mbox{higher derivative terms}.
\end{eqnarray*}
which combined with (\ref{Boris}) gives us
\begin{equation}
H_b = \int \left(\hat{E}^2   + F^2 \right) + \co(\tau_1^{{1/2p}}) + \mbox{higher derivative terms}
\label{Hbb}
\end{equation}
where the higher orders in $\tau_1^{{1/2p}}$ are computed via the integral kernel $K_{f,A}$. We recognize the first term in (\ref{Hbb}) as the Hamilton operator of a Yang-Mills theory.


\begin{remark}
Note that the inclusion of the complex phase to the ground state, that involved the Chern-Simons term in (\ref{first}), is essential for the Hamilton operator (\ref{Hbb}) to emerge. It is interesting that a topological term is required for a Yang-Mills type theory to emerge from this formalism.
\end{remark}

From the perspective of this article, instead of the formal operators $\hat{E}({\bf m})$ in formula (\ref{vladp}) it is more natural to consider the operators $\hat{E}_A({\bf m})$.
In this case the commutation relations also change
$$
\left( \left[ \hat{E}_A({\bf m}_2),\hat{A}({\bf m}_1) \right] \eta\right)(A)  = K_{f,A}({\bf m}_2,{\bf m}_1) \eta(A)
$$
where to lowest order in $\tau_1$ gives us the canonical commutation relations
\begin{equation}
\left( \left[ \hat{E}_A ({\bf m}_2),\hat{A}({\bf m}_1) \right]\eta\right)(A)  = \d({\bf m}_2-{\bf m}_1) \eta(A) + \co(\tau_1^{{1/2p}}).
\label{Hbbb}
\end{equation}
Combining (\ref{Hbbb}) with (\ref{Hbb}) shows that in the limit $\tau_1\rightarrow 0$ our formalism coincides with that of a Yang-Mills theory on a curved background and which involves higher derivative interactions. The expansions $\hat{E}({\bf m})=\sum_i \xi_i({\bf m})\frac{\pa}{\pa \xi_i}$ and $A=\sum_i x_i \xi_i({\bf m})$ correspond to the plane wave expansions of the conjugate fields, which are used in ordinary quantum field theory on a flat manifold.

\begin{remark}
Note that relation (\ref{Hbbb}) implies that causality will be affected at the scale of $\tau_1^{{1/2p}}$. This suggest that $\tau_1^{{1/2p}}$ should be interpreted as the Planck length.
\end{remark}

\subsection{The fermionic sector}

Let us now turn to the fermionic sector given by the Hamiltonian $H_f$ in (\ref{Hf}). If we apply a second derivation to equation (\ref{godhavn}) we obtain
$$
\frac{\pa^2 CS}{\pa \xi_i \pa \xi_j} =  2 \int_M \mbox{Tr}\left(\xi_i\wedge \nabla^A \xi_j \right) ,      
$$
which gives us
\begin{equation}
H_f  = \mathrm{i}\int \left(  {\phi}  \nabla^A \bar{\phi}\right) + \mbox{higher order terms} \quad (\tau_1=0)
\label{helle}
\end{equation}
where $\bar{\phi}=\sum_i\bar{c}_i\xi_i$ and ${\phi}=\sum_i {c}_i\xi_i$ and where $\nabla^A$ is again the covariant derivative $\nabla^A=d+A$. Also, here "higher derivative terms" include both terms coming from $S(A)$ and the term coming from the single derivative of the Chern-Simons functional in (\ref{Hf}). The expression in (\ref{helle}) looks similar to a Dirac Hamiltonian of conjugate fermionic fields $\phi$ and $\bar{\phi}$. Note, however,  that these fermionic fields due to the Clifford relations in (\ref{mangec}) satisfy the relations
\begin{eqnarray}
\left\{ \bar{\phi} ({\bf m}_1), \bar{\phi}({\bf m}_2)  \right\} &=& \d({\bf m}_1-{\bf m}_2) + \co(\tau_1^{{1/2p}}),
  \nn\\
\left\{ {\phi} ({\bf m}_1), {\phi}({\bf m}_2)  \right\} &=& \d({\bf m}_1-{\bf m}_2) + \co(\tau_1^{{1/2p}}),
  \nn\\
\left\{ \bar{\phi} ({\bf m}_1), {\phi}({\bf m}_2)  \right\} &=& 0,
\end{eqnarray}
which are \underline{not} the canonical commutation relations of a fermionic field and its conjugate but rather a twisted version hereof. 

\begin{remark}
Note that the term written in (\ref{helle}) is topological. Note also that the complete Hamilton $H_b+H_f$ resembles a model, which Witten formulated in his paper \cite{Witten:1982im} on Morse theory.
\end{remark}

The fermions, which emerge from this formalism, are vectors. This issue, which appears to be at odds with the spin-statistics theorem and thus with special relativity, will be discussed in section \ref{discussion}.

\section{The case with a Levi-Civita connection on $\ca$}
\label{levi}

In section \ref{secdirac} when we constructed both the Dirac and the Bott-Dirac operator we indirectly assumed the existence of a global trivialization, which we used to make sense of $\frac{\pa}{\pa\xi}$. In this section we will discuss the more canonical choice, which is to use the Levi-Civita connection.

In this case the Bott-Dirac operator (\ref{hmm}) will have the form
$$
B^{(\pm)} \eta =\sum_i (c_{i} \nabla_{\xi_i}\eta + \bar{c}_{i} d^{(\pm)}_i \eta) .
$$
where once more
$$
d^{(\pm)}_i = \frac{\partial S^{(\pm)} }{\partial \xi_i }
$$
and where 
$$
\nabla_{\xi_i}= \frac{\pa}{\pa\xi_i}+ \oo_{\xi_i}
$$
with $\oo_{\xi_i}$ being the unique Levi-Civita connection in $T\ca$ that preserves the metric (\ref{inner}) and has vanishing torsion. Note that $B^{(\pm)} \Psi^{(\pm)}=0$ still holds since $\Psi^{(\pm)}$ is a scalar in $T\ca$.


With $\oo$ we can write down the curvature 
$$
F_{ij}(\oo)= \frac{\pa}{\pa \xi_i} \oo_{\xi_j} - \frac{\pa}{\pa \xi_j} \oo_{\xi_i} + [\oo_{\xi_j},\oo_{\xi_j}]
$$
on $\ca$, which will then show up in the square of the Bott-Dirac operator via the general Bochner identity.
Alternatively, we can consider computing the spectral action \cite{Chamseddine:1996zu,Chamseddine:2008zj} of the Bott-Dirac operator. 

It is beyond the scope of this paper to investigate these options. The point, which we wish to make here, is simply that in a local limit a gravitational theory on $\ca$, as we have demonstrated in these pages, will look like a quantum field theory on the manifold $M$.


\section{Discussion}
\label{discussion}

In this paper we have presented a framework of non-perturbative quantum field theory in which a candidate for a quantum field theory, that involves a quantized gauge field coupled to a fermionic field on a curved background, emerges from a geometrical construction over a configuration space of gauge connections.
This construction raises a number of critical questions concerning both technical aspects as well as aspects of interpretation. 

Clearly the most important question is that of existence. We know that the representation of the $\mathbf{QHD}(M)$ algebra exist in the non-covariant case, i.e. the case where the Sobolev norm on $\OO^1(M,\mathfrak{g})$ is not covariant, but we do not know whether this result holds when we require the gauge symmetry to be preserved. We expect, however, that this will be the case: the covariant Sobolev norm still functions as an ultra-violet regulator and thus it seems very likely that the power of the covariant Hodge-Laplace operator can be chosen sufficiently high to ensure convergence. 

The second critical question is concerned with the fermionic sector. As it stands now it needs improvements. The fermions have the wrong spin-statistics and the fermionic sector of the Hamilton operator does not have the right form. Clearly something is missing. 

The basic idea is that it should be possible to interpret the fermionic sector in a quantum field theory in terms of an infinite-dimensional Clifford algebra associated with a geometrical structure -- such as the Bott-Dirac operator -- on the configuration space of connections. But for this idea to be physically feasible we need to find a way to realize it using a CAR algebra of spin-half fermions. 
There are several avenues to explore here. First, it may be that we have not fully understood the representations of the CAR algebra; it could be that there exist spin representations also in the infinite-dimensional case. Whether such representations can be reached through projective and inductive techniques is, however, not clear.
Secondly, it may be that the construction of the Bott-Dirac operator as it stands is too crude and that it is possible to link degrees of freedom in $T\ca$ directly to a CAR algebra build from spinors in a more refined manner.

A related question is that of Lorentz invariance. It is not clear that our construction preserves the Lorentz symmetry, nor is it clear that such an invariance is achievable. This of course also concerns causality. More work is required to determine to what extend the Lorentz symmetry is broken, but one possibility is that it will be modified with a scale transformation, the result being a Lorentz violation at the Planck scale, something that may very well be within experimental bounds. It is plausible that such considerations will eventually put limits on how the metric on $T\ca$ can be constructed, i.e. precisely how the covariant Hodge-Laplace operator is employed. 

These technical issues aside there is also the question concerning the physical interpretation of our formalism. This hinges on the choice of the configuration space of connections. There appear to be two options: either we are dealing with a Yang-Mills type of theory coupled to fermions on a curved background, in which case the choice of configuration space is essentially open; it merely reflects what kind of Yang-Mills theory we wish to formulate. The other option is to interpret the formalism as a gravitational theory and the configuration space as a space of spin-connections, possible Ashtekar connections. In this case one needs to understand what role the background metric plays in relationship to the conjugate variables, which would then also be gravitational.

The basic idea that the ultra-violet regularisation should be considered as physical rather than a computational tool, is in our opinion one worth considering in a broader spectrum of settings. It necessitates that the regularisation is made covariant, as explored in this paper, which places it in some proxemity to non-local quantum field theory. Particularly, it is an interesting question whether it could be applied to non-perturbative Yang-Mills theory on curved backgrounds, both in Hamiltonian and Lagrangian (Euclidean) approaches.

The formalism that we present depends on a metric on the underlying manifold. As it stands now this metric does not appear to be dynamical, i.e. we do not appear to be able to derive also the Hamilton for this gravitational sector. It is possible that this will be contained in the final -- and largely undeveloped -- version of our formalism where we instead of a global trivialisation of $T\ca$ use a Levi-Civita connection in $T\ca$. More analysis is required to determine whether this is the case. 
Another, related issue is what physical role geometrical invariants on $\ca$ and $T\ca$ may play.


Finally, the formalism that we present employs the toolbox of non-commutative geometry. The $\mathbf{QHD}(M)$ algebra is non-commutative and it interacts with a Bott-Dirac operator. These are ingredients familiar to the non-commutative geometry aficionado. We believe that a more throrough application of the noncommutative geometry-toolbox is possible and desirable.

\vspace{1cm}
\noindent{\bf\large Acknowledgements}\\

\noindent
JMG would like to express his gratitude towards Ilyas Khan, United Kingdom, and towards the engineering company Regnestuen Haukohl \& K\o ppen, Denmark, for their generous financial support. JMG would also like to express his gratitude towards the following sponsors:  Ria Blanken, Niels Peter Dahl, Simon Kitson, Rita and Hans-J\o rgen Mogensen, Tero Pulkkinen and Christopher Skak for their financial support, as well as all the backers of the 2016 Indiegogo crowdfunding campaign, that has enabled this work. Finally, JMG would like to thank the mathematical Institute at the Leibniz University in Hannover for kind hospitality during numerous visits.\\

\end{document}